# The Rhythm of Tai Chi:
# Revitalizing Cultural Heritage in Virtual Reality through Interactive Visuals

Xianghan Wang*
*Integrated Digital Media*
*New York University*
New York, the United States
xw2264@nyu.edu

*Abstract*—The Rhythm of Tai Chi reinterprets the ancient Chinese martial art as a dynamic, interactive virtual reality (VR) experience. By leveraging computer vision and multimedia technologies, the project transforms Tai Chi's philosophy and movements into an immersive digital form. Real-time motion tracking captures user gestures, while visual feedback systems simulate the flow of Qi, enabling an intuitive and engaging practice environment. Beyond technological innovation, this work bridges traditional Chinese culture and modern audiences. It offers a global platform—accessible even to those unfamiliar with Tai Chi—to explore its cultural significance, connections to balance, health, and mindfulness. Serving as both a preservation tool and an educational resource, The Rhythm of Tai Chi revitalizes this heritage for the digital age.

## I. INTRODUCTION

In a rapidly evolving digital world, the preservation and transmission of cultural heritage face significant challenges. Traditional cultures like Tai Chi, a centuries-old Chinese martial art, hold profound cultural, philosophical, and wellness values. However, Tai Chi struggles to reach contemporary audiences, particularly younger generations. While CNN highlighted millennials' growing curiosity about Tai Chi, practical barriers such as difficulty finding instructors, understanding the cultural context, and connecting to communities limit their participation [1]. To prevent this cultural heritage from fading, innovative approaches are urgently needed.

Immersive technologies such as VR and computer vision offer transformative potential for cultural preservation. By creating dynamic, interactive environments, they make traditional practices like Tai Chi more accessible and engaging to modern audiences [2]. Real-time motion tracking and gesture recognition enable physical participation, while multimedia visualization helps convey abstract concepts such as the flow of Qi. The Rhythm of Tai Chi leverages these technologies to reinterpret Tai Chi for the digital age. Through an interactive VR platform, the system transforms physical gestures into real-time visualizations of Qi, linking movement to cultural philosophy and promoting intuitive understanding. This approach enhances immersion, cultural depth, and accessibility, transcending geographical and pedagogical barriers. This paper explores how immersive technologies can revitalize traditional practices by combining technical innovation with cultural preservation. We demonstrate how VR and computer vision enhance engagement through interactivity, visualize intangible cultural concepts, and foster connections between ancient wisdom and modern lifestyles. Our work provides a scalable model for safeguarding intangible heritage through technology.

## II. REINTERPRETING TAI CHI THROUGH VIRTUAL REALITY

Tai Chi, a centuries-old Chinese martial art, represents more than just physical movement—it embodies a profound cultural philosophy rooted in balance, harmony, and the flow of Qi (life energy). Tai Chi has long been known for its health benefits, including improving cardiopulmonary function, calming the nervous system, and alleviating discomfort in the lumbar and cervical spine, providing relief from modern health issues [3]. Furthermore, Tai Chi and Traditional Chinese Medicine have been recognized globally for their therapeutic effects, with organizations like the World Health Organization (WHO) promoting their practice for overall well-being [4].

A major challenge in traditional Tai Chi teaching is the abstract concept of Qi. While experienced practitioners are able to sense and regulate Qi through physical coordination and mental concentration, beginners, especially those unfamiliar with Chinese culture, often find it difficult to intuitively observe Qi and understand its dynamic movement within the body. The flow of Qi through meridians (jingluo) connects breath, movement, and emotion into a unified system, which is central to the concepts of Traditional Chinese Medicine [5]. However, conventional teaching methods, such as verbal instruction or static diagrams, are often insufficient in effectively conveying this dynamic flow.

The Rhythm of Tai Chi addresses this challenge by using virtual reality to transform Tai Chi's movements and philosophy into a dynamic, interactive experience. Integrating computer vision and multimedia, the system visualizes the flow of Qi, making this invisible energy more tangible and accessible. Users engage not only with physical movement but also with the cultural and philosophical meaning behind it, bridging traditional heritage with modern technology.

## III. TECHNICAL APPROACH

The technical implementation of The Rhythm of Tai Chi combines VR, motion tracking, and computer vision to create an immersive and interactive experience that brings Tai Chi into the digital age. Developed in Unreal Engine 4.27 and deployed on the Oculus Quest 2, the system offers a wireless, portable platform that preserves the openness essential to Tai Chi practice.

### A. Core System Architecture

Building on the foundation introduced above, the system integrates the rendering and physics capabilities of Unreal Engine with the wireless portability of the Quest 2 to create an environment tailored to Tai Chi's meditative flow. VR headsets allow users to move without physical restraints, preserving the openness and full-body engagement required of Tai Chi.

Unreal Engine supports real-time rendering, motion tracking, particle interactions, and environmental feedback. The Niagara particle system is used to simulate dynamic environmental effects such as falling snow, swirling sand, and glowing Qi trails. These features allow the virtual world to respond to users' movements in real time, creating an environment that feels immersive and reactive. This integration provides the foundation for a fluid, embodied experience that blends cultural immersion with physical engagement.

### B. Design Philosophy and Interaction Flow

The design of The Rhythm of Tai Chi is grounded in Tai Chi philosophy, with a focus on balance, breath, flow, and the unity of mind, body, and nature. Rather than treating Tai Chi as a goal-driven game, the system emphasizes experience over performance. This approach intentionally avoids competitive features like scoring or ranking, aligning instead with Tai Chi's meditative and restorative nature.

A mind map was created to structure the Tai Chi VR experience, organizing key elements across space, time, interaction, and progression. It ensures that Tai Chi's physical movements flow naturally within the virtual environment, with each component—motion capture, visualization, environment, and guidance—reinforcing its core cultural values and the connection between mind, body, and nature [6].

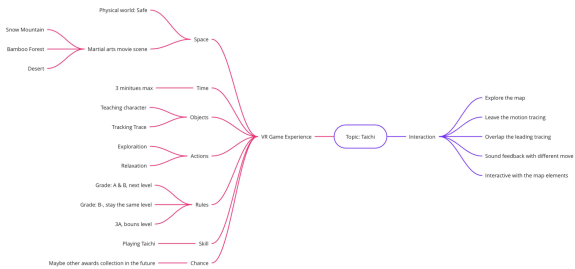

Fig. 1. Mind map

The interaction flow follows a continuous cycle in which users perform Tai Chi movements, their gestures are captured, the system visualizes Qi, and the environment responds accordingly. This feedback loop enables users to see the real-time effects of their actions and intuitively sense the connection between internal energy and the surrounding world. The result is an immersive experience where movement, meaning, and mindfulness are deeply integrated.

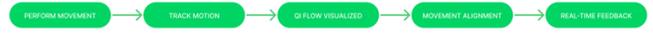

Fig. 2. Flowcharts of the Qi visualization process

### C. Real-Time Motion Tracking and Interactive Visual Feedback Systems

The Rhythm of Tai Chi integrates real-time motion tracking and interactive visual feedback systems to create an immersive and intuitive experience that bridges the gap between traditional cultural practices and modern technology. One of the core elements of the system is motion tracking, which captures the user's physical movements and translates them into dynamic actions in the virtual environment. Unreal Engine's powerful motion tracking system, combined with the precision of Oculus Quest 2's controller input, enables real-time recognition of gestures and alignment, critical to replicating the fluid movements of Tai Chi.

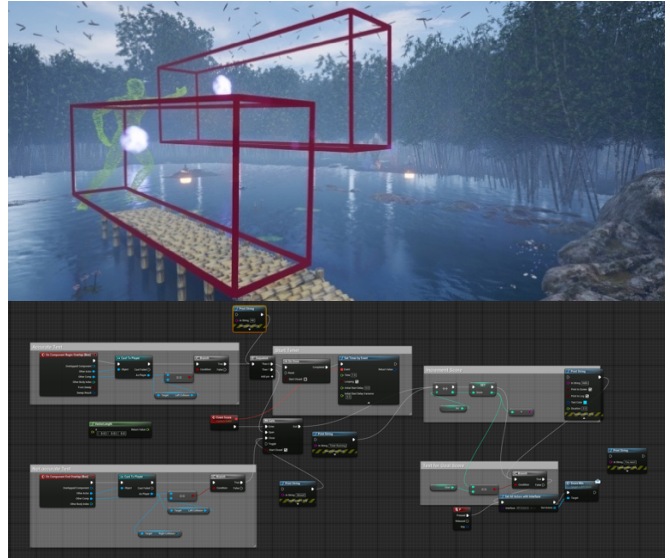

Fig. 3. Motion tracking setup in Unreal Engine

As discussed previously, one significant challenge in learning Tai Chi is the abstract nature of its fundamental concept, Qi. To address this, The Rhythm of Tai Chi employs a specialized visualization system that combines motion tracking data with interactive visual feedback. This system transforms the intangible concept of Qi into a tangible, observable phenomenon, vividly illustrating the flow of life energy as users perform Tai Chi movements. In real-time, users witness their Qi represented as a dynamic, luminous trail, visually reinforcing their experiential and educational understanding of the practice.

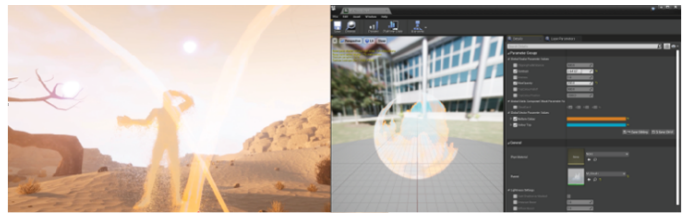

Fig. 4. Visualization of Qi's Flow

When users perform movements with calm focus and fluid alignment, their Qi manifests as a soft, glowing flow, like ink swirling in water or light diffusing in mist. The visual trail responds to the user's movement, thickening, brightening, and becoming more harmonious as the Tai Chi movements improve in alignment and smoothness. If the user's movements are imbalanced or erratic, the Qi trail becomes unstable, breaking apart or flickering, creating a powerful feedback loop that helps the user intuitively understand the corrections needed. Similar approaches in immersive systems have shown that posture-specific feedback can effectively guide users toward more accurate and intentional movement patterns in VR environments [7]. In the context of Tai Chi, this system offers a more immediate and visually accessible way to perceive the flow of Qi, allowing users to see, rather than just imagine, the internal energy dynamic of their practice.

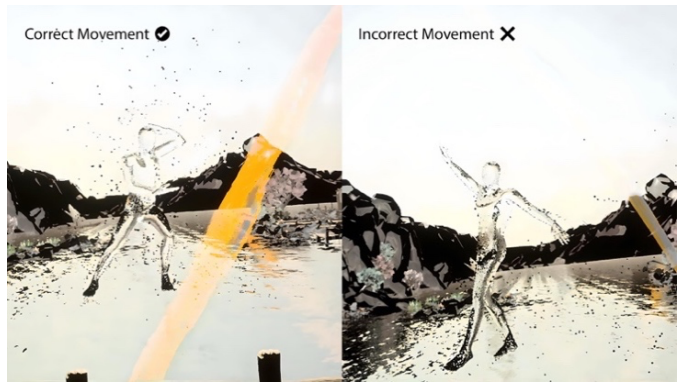

Fig. 5. Visualization feedback of Qi for both correct and incorrect movement

The system uses the visualization of Qi's thickness and flow as a means to assess the accuracy of the user's movements, which is aligned with Tai Chi's inherent qualities. Tai Chi is traditionally practiced for relaxation and stress relief, not for competition or scoring. This distinguishes the project from typical game mechanics, which focus on achieving high scores. Instead, The Rhythm of Tai Chi is intentionally designed around reflective practice, internal harmony, and continuous self-improvement. The real-time visualization dynamically reflects the user's internal state, offering nuanced feedback on subtle improvements in form, balance, and the smoothness of transitions between movements.

By avoiding the conventional gamification approach, the project redefines the experience as an immersive experience that allows users to engage with Tai Chi in its truest form. The real-time visualization of Qi becomes a tool for users to understand the fluidity and harmony of their movements, connecting them to the cultural and philosophical essence of Tai Chi. This immersive experience emphasizes the practice's meditative and restorative qualities, rather than competitive achievement, making it an accessible and mindful learning process.

### D. Immersive Environment Design

To attract more young people to experience the charm of Tai Chi, especially since it has long been considered a slow exercise for the elderly, The Rhythm of Tai Chi draws heavy inspiration from martial arts movies, incorporating interactive natural elements like snow-capped mountains, bamboo forests, deserts, and ink paintings to enhance immersion while reinforcing Tai Chi's martial arts heritage. Each environment is not just a backdrop but an integral part of the experience, where the user's actions directly interact with and impact the world around them.

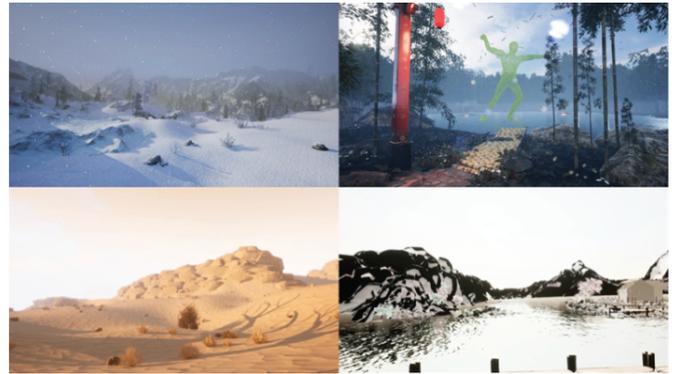

Fig. 6. Level design of the Rhythm of Tai Chi

For instance, users' movements leave footprints in the snow and sand, creating a tangible visual record of their practice. Grass gently bends underfoot, reflecting the fluid and continuous nature of Tai Chi movements, and gradually returns upright, mirroring the user's sense of balance and flow. Dynamic weather conditions shift in response to the user's internal state—calm, balanced movements bring clearer skies, while imbalanced gestures trigger storms or gusts of wind. In the bamboo forest setting, subtle environmental reactions such as bending bamboo stalks and rustling leaves emphasize Tai Chi's principle of harmony between movement and stillness. Similarly, in the desert environment, visual effects such as heat distortion and swirling dust dynamically follow the user's motions, vividly illustrating the flow and intensity of Qi, and reinforcing the profound interconnectedness between human action and nature inherent in Tai Chi.

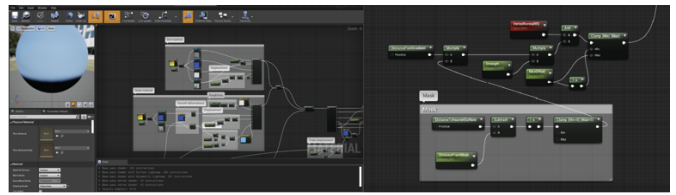

Fig. 7. Examples of Interactive Snow and Grass Blueprints

The project also incorporates a dynamic weather system that not only helps users gain a deeper understanding of their Qi flow but also reinforces the meditative and reflective nature of Tai Chi, where internal harmony influences the external world. After completing a Tai Chi session, elements such as wind, snow, sand, and fog gradually clear, symbolizing the restoration of balance and harmony. The dynamic weather system enhances the immersive experience by adapting to the user's internal balance and Qi flow. When users move with calm focus and fluid alignment, the environment reacts positively, with clear skies and gentle winds, representing harmony and stability. However, when the user's movements become imbalanced or erratic, the environment reflects this instability with turbulent weather—such as storms, gusts of wind, or shifting sandstorms. This dynamic feedback serves as a visual metaphor for the flow of Qi, providing users with an immediate representation of their practice's alignment or imbalance.

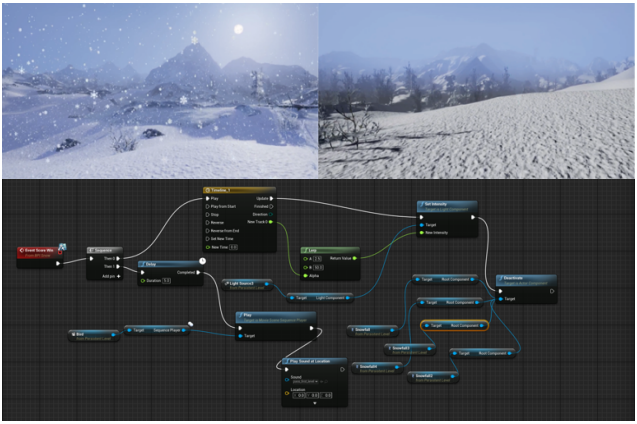

Fig. 8. Examples of dynamic weather system

### E. Particle-Based Mentors

The particle-based mentors, designed using Unreal Engine's Niagara system, take humanoid forms inspired by natural elements such as wind, water, and snow, reflecting Tai Chi's philosophy of neutrality. These mentors are humanoid in form but gender-neutral, ensuring that users focus on the Qi (life energy) and their connection with nature rather than any human identity. By using elemental particles to form human-like shapes, these mentors reflect the fluidity and impermanence of nature, which is central to Tai Chi's philosophy. This design choice maintains the neutrality of Tai Chi, allowing the user to engage with the practice without the distraction of human representations.

The particle mentors act as subtle guides throughout the user's Tai Chi practice, providing assistance without imposing human characteristics. Their fluid, ever-changing forms mirror the impermanence of nature, reinforcing the connection between the user and the Qi. These mentors, while humanoid, focus on guiding the user in their understanding of Tai Chi's movements, allowing them to better align with nature's forces. This approach ensures that the user's attention remains on the movement of Qi, preserving the cultural and philosophical essence of Tai Chi, and enhancing the overall immersive experience.

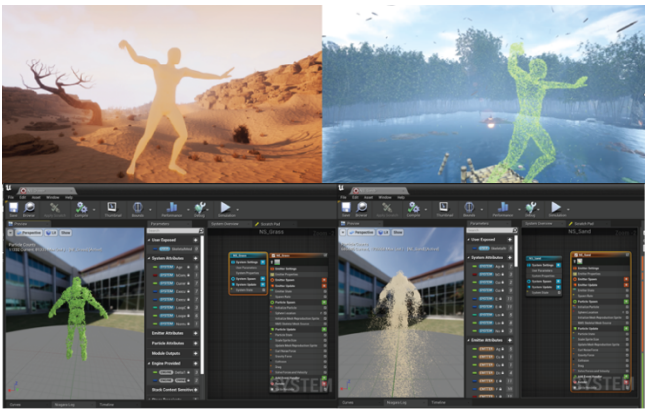

Fig. 9. Examples of particle-based mentors

### F. User Engagement

The Rhythm of Tai Chi addresses three key challenges in the modern transmission of traditional Tai Chi. First, it counters the perception of Tai Chi as "slow and boring," especially among younger audiences. Since Tai Chi requires long-term practice to yield noticeable results, it often fails to retain users with shorter attention spans. The project uses VR, visual feedback, and light gamification to spark curiosity and make Tai Chi practice more engaging [8]. As discussed earlier, immersive and interactive elements are particularly effective in capturing attention and increasing user retention.

Second, the system transforms sustained practice into a series of short, tangible achievements by integrating real-time motion feedback and Qi visualization. This immediate feedback loop makes the experience more dynamic and enjoyable, helping users stay motivated through continuous visual response [9]. By seeing the flow of Qi as they move, users gain a clearer sense of progress, bridging the gap between philosophical understanding and physical execution.

Finally, VR-assisted practice significantly enhances user engagement. Research indicates that VR increases the frequency, duration, and intensity of exercise, with users showing improvements in functional ability by 31% and muscular strength by 24%. These findings confirm that VR not only encourages more frequent practice but also boosts user motivation and satisfaction [10]. Therefore, VR presents an innovative approach to practicing Tai Chi, making it more accessible while amplifying its physical and mental health benefits.

### G. Preliminary Results

To assess how these engagement strategies translate into real-world impact, a small-scale user test was conducted with approximately 30 participants. Over 90% reported that they had never practiced Tai Chi before but became interested in learning more after the experience. Around 90% also said they could better sense the flow of Qi, and 95% enjoyed the immersive environment. Many users expressed a desire for future content and indicated they would consider paying for a fully developed version of the system.

TABLE I. PARTICIPANT FEEDBACK SUMMARY (N ≈ 30)

| Feedback Item | Response (%) |
| --- | --- |
| Never practiced Tai Chi before | 90%+ |
| Interested in learning after experience | 70%+ |
| Better sensed the flow of Qi | ~90% |
| Enjoyed the immersive environment | 95% |

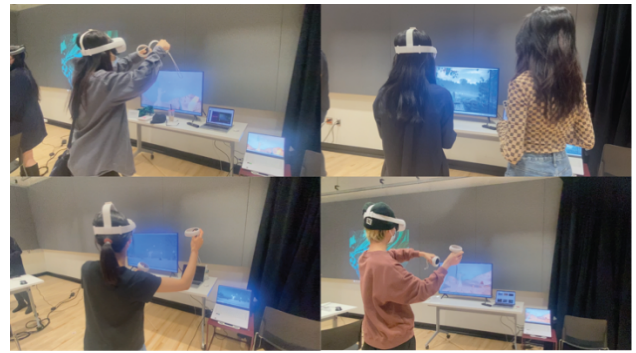

Fig. 10. User testing in VR with real-time feedback

## IV. Future Directions

VR technology has shown strong potential in cultural heritage preservation, with notable applications such as Dunhuang VR, Notre-Dame de Paris VR Experience, and Microsoft's mixed reality museum in Kyoto, which offers an immersive way to view centuries-old Japanese artwork through holographic layering and spatial sound design [11]. While these projects excel at providing immersive access to visual artifacts and historical environments, they are primarily centered on passive exploration or observation.

In contrast to observation-based VR projects, The Rhythm of Tai Chi emphasizes embodied interaction, allowing users to physically perform rather than passively view cultural practices. The system translates Tai Chi's core philosophies into a real-time sensory feedback loop using motion tracking and interactive visualization. This approach enables users to internalize cultural meaning through action, not just perception. This model can be extended to other forms of intangible cultural heritage, including ritual performance, embodied knowledge, and multisensory traditions. By combining immersive environments with real-time feedback, the approach provides a scalable framework for preserving diverse cultural expressions.

Despite its potential, The Rhythm of Tai Chi faces several scalability challenges. Content development remains resource-intensive, as accurately modeling culturally specific environments, gestures, and feedback systems requires interdisciplinary collaboration and time-consuming design. Broader adoption is also limited by hardware accessibility and the need for cultural adaptation across different regions. Addressing these challenges will require close collaboration with cultural experts to ensure authenticity, and with technologists to develop more adaptable, multimodal interaction systems. These efforts could be supported by a shared infrastructure such as a global cultural heritage database, enabling scalable and culturally responsive development.

To that end, future work could focus on developing a global digital cultural heritage database to preserve endangered traditions. This database would centralize cultural assets, particularly those challenging to document conventionally. Collaboration with cultural experts, anthropologists, local communities, and technologists would ensure cultural authenticity and adapt traditional practices into diverse digital formats. Methods such as 3D artifact scanning, motion capture, and virtual environment reconstruction would facilitate content creation. Rather than offering a single mode of preservation, this infrastructure would enable multiple forms of interaction, like educational, experiential, performative, ensuring intangible heritage is both protected and reactivated in modern contexts. Artificial intelligence could further enhance this system by integrating and reconstructing heritage content, and combined with virtual reality, tailor personalized, adaptive conservation experiences based on users' cultural backgrounds, physical abilities, or learning preferences [12]. In this collaborative framework, contributions from museums, researchers, artists, and developers would ensure not only technical quality and scalability, but also cultural diversity and ethical integrity in what is preserved and how it is shared.

## V. Conclusion

In conclusion, The Rhythm of Tai Chi demonstrates how immersive technologies such as VR and computer vision can reframe traditional cultural practices into interactive, embodied experiences. By visualizing Qi and translating Tai Chi's philosophical principles into real-time feedback, the project bridges the gap between intangible heritage and modern digital interaction. It offers a scalable, experience-driven model for cultural preservation that is both engaging and educational. This work contributes not only to the design of motion-based heritage systems, but also to the broader discourse on how immersive tools can support meaningful cultural continuity in the digital age.

### Acknowledgment

I would like to thank everyone who supported me throughout this project. My deepest gratitude goes to my advisor, Todd Bryant, for his invaluable guidance and encouragement. Special thanks to my family for their continuous support and for inspiring my passion and determination through the cultural heritage they shared with me.